# Nature of charge density wave in kagome metal ScV$_6$Sn$_6$

Seongyong Lee[1,2], Choongjae Won[1], Jimin Kim[1,2], Jonggyu Yoo[1,2], Sudong Park[1,2], Jonathan Denlinger[3], Chris Jozwiak[3], Aaron Bostwick[3], Eli Rotenberg[3], Riccardo Comin[4†], Mingu Kang[1,4†] & Jae-Hoon Park[1,2†]

[1]Max Planck POSTECH/Korea Research Initiative, Center for Complex Phase of Materials, Pohang 790-784, Republic of Korea.

[2]Department of Physics, Pohang University of Science and Technology, Pohang 790-784, Republic of Korea.

[3]Advanced Light Source, E. O. Lawrence Berkeley National Laboratory, Berkeley, California 94720, USA.

[4]Department of Physics, Massachusetts Institute of Technology, Cambridge, Massachusetts 02139, USA.

†Correspondence should be addressed to rcomin@mit.edu, iordia@mit.edu, jhp@postech.ac.kr

**Kagome lattice materials offer a fertile ground to discover novel quantum phases of matter, ranging from unconventional superconductivity and quantum spin liquids to charge orders of various profiles. However, understanding the genuine origin of the quantum phases in kagome materials is often challenging, owing to the intertwined atomic, electronic, and structural degrees of freedom. Here, we combine angle-resolved photoemission spectroscopy, phonon mode calculation, and chemical doping to elucidate the driving mechanism of the $\sqrt{3}\times\sqrt{3}$ charge order in a newly discovered kagome metal ScV$_6$Sn$_6$. In contrast to the case of the archetype kagome system $A$V$_3$Sb$_5$ ($A$= K, Rb, Cs), the van Hove singularities in ScV$_6$Sn$_6$ remain intact across the charge order transition, indicating a marginal role of the electronic instability from the V kagome lattice. Instead, we identified a three-dimensional band with dominant planar Sn character opening a large charge order gap of 260 meV and strongly reconstructing the Fermi surface. Our complementary phonon dispersion calculations further emphasize the role of the structural components other than the V kagome lattice by revealing the unstable planar Sn and Sc phonon modes associated to the $\sqrt{3}\times\sqrt{3}$ phase. Finally, in the constructed phase diagram of Sc(V$_{1-x}$Cr$_x$)$_6$Sn$_6$, the charge order remains robust in a wide doping range $x \approx 0 \sim 0.10$ against the Fermi level shift up to $\approx$ 120 meV, further making the electronic scenarios such as Fermi surface or saddle point nesting unlikely. Our multimodal investigations demonstrate that the physics of ScV$_6$Sn$_6$ is fundamentally different from the canonical kagome metal $A$V$_3$Sb$_5$, uncovering a new mechanism to induce symmetry-breaking phase transition in kagome lattice materials.**

Strongly correlated systems and topological materials are two different research areas in condensed matter physics, while exotic electronic phenomena often appear at their intersection.[1] Kagome lattice, a two-dimensional network of corner-sharing triangles (Fig. 1a), naturally lies at this intersection thanks to the unique symmetry-protected electronic structure composed of Dirac fermions at K, van Hove singularities (vHS) at M, and a flat band across the whole Brillouin zone (Fig. 1a). On one hand, the diverging density of states at the van Hove singularities and flat band fillings can promote various correlated many-body ground states.[2–6] On the other hand, the linear band crossing at K and the quadratic band touching degeneracy at $\Gamma$ can be a singular source of Berry curvature and nontrivial topology.[7–9] Accordingly, the kagome lattice materials offer a promising opportunity to discover novel electronic phenomena at the confluence of correlation and topology, and have attracted significant research interests during the past few years.[10–17]

A family of $A$V$_3$Sb$_5$ ($A$= K, Rb, Cs, Fig. 1b) represents an archetype kagome system hosting a rich series of emergent electronic orders, including the 2×2 charge order,[17] 1×4 stripe order,[18] electronic nematicity,[19] superconductivity,[17] and pair density waves.[20] The electronic instabilities associated with the vHSs of the V kagome lattice have been proposed as the origin behind these rich behaviors of $A$V$_3$Sb$_5$.[21–23] Among these, the 2×2 charge order (Fig. 1g) exhibits many unconventional characteristics and has been considered a key to understand the exotic physics of $A$V$_3$Sb$_5$. For example, the 2×2 charge order in $A$V$_3$Sb$_5$ may accompany an imaginary conjugate component called chiral flux order,[24,25] which may explain the spontaneous time-reversal symmetry breaking[26,27] and anomalous Hall conductivity[28] observed in $A$V$_3$Sb$_5$ without explicit magnetism. Also, an intricate competition between the 2×2 charge order and superconductivity gives rise to the multiple superconducting domes in the phase diagrams of CsV$_3$Sb$_5$.[29–31] In this context, understanding the nature of unconventional charge orders in kagome lattice materials is of fundamental importance in this emerging research field.

Meanwhile, the 2×2 charge order is not the only form of the charge order proposed in the kagome lattice: early theories predicted more diverse forms of charge order to appear at 1/3, 2/3, and vHS fillings, including not only the 2×2 charge bond order (Fig. 1g) but also the 1×1, 2×1, and $\sqrt{3}\times\sqrt{3}$ charge density waves (Fig. 1h-j).[32–35] The latter phases are distinguished from the 2×2 charge bond order in $A$V$_3$Sb$_5$ by the charge disproportionation at each lattice site, which reflects the manifestation of long-range Coulomb interaction. Intriguingly, such charge disproportionation phases in the frustrated kagome lattice geometry have been predicted to host a fractional charge excitation e/2,[33] analogous to the fractional spin excitations in kagome quantum spin liquids.[36] In this respect, exploring a new kagome system hosting diverse forms of charge order is highly desired, yet has been missing so far.

To this end, we turn our attention to the newly discovered kagome compound ScV$_6$Sn$_6$ hosting the novel $\sqrt{3}\times\sqrt{3}$ charge order below $T_{CO} \approx 92$ K (Fig. 1j).[37] The ScV$_6$Sn$_6$ belongs to the large family of HfFe$_6$Ge$_6$-type '166' kagome metals (Fig. 1c) with a prospect to tune the charge order by broad chemical substitutions.[16] However, the origin and nature of the $\sqrt{3}\times\sqrt{3}$ charge order in ScV$_6$Sn$_6$ have remained to be understood. On one side, the $\sqrt{3}\times\sqrt{3}$ charge order may be a consequence of the intrinsic electronic instability of the kagome lattice as predicted from the extended Hubbard model since the early 2010s (Fig. 1j).[32–34] The ScV$_6$Sn$_6$ shares the partially filled V kagome lattice with the $A$V$_3$Sb$_5$, so it is tempting to suggest that the same vHS instability of $A$V$_3$Sb$_5$ also contributes to the charge order in ScV$_6$Sn$_6$. On the other side, the X-ray refinement of the charge order structure revealed the dominant displacement of the Sc and Sn atoms, while the displacement within the V kagome lattice is marginal.[37] Moreover, the $\sqrt{3}\times\sqrt{3}$ charge order is not generally observed in $R$V$_6$Sn$_6$ series ($R$ = Sc, Y, and rare earth elements), suggesting that extrinsic factors specific to ScV$_6$Sn$_6$ may play a role.

In this work, we established the origin of the $\sqrt{3}\times\sqrt{3}$ charge order in ScV$_6$Sn$_6$ by comprehensively mapping its electronic structure, phonon dispersion, and phase diagram. Our multimodal approaches coherently point toward that the $\sqrt{3}\times\sqrt{3}$ charge order in ScV$_6$Sn$_6$ is tied to the structural components other than the V kagome lattice and is thus fundamentally different from the 2×2 charge order in $A$V$_3$Sb$_5$ originating from the intrinsic electronic instability of V kagome plane.

We start with the basic characterizations of ScV$_6$Sn$_6$. Our transport measurements revealed a sudden change in electrical resistivity around $T_{CO} \approx 92$ K, signaling a symmetry-breaking phase transition (Fig. 1f). X-ray diffraction measurements detected commensurate superlattice peaks in the low-temperature phase consistent with the $\sqrt{3}\times\sqrt{3}$ charge ordering (Fig. 1d). Both the abrupt drop in the diffraction peak intensity at $T_{CO}$ (Fig. 1e) and the small thermal hysteresis in resistivity (Fig. 1f) are indicative of the first-order nature of the transition. Overall, our transport and diffraction characterizations of ScV$_6$Sn$_6$ are in close agreement with the original report.[37]

Before discussing the detailed electronic structure of ScV$_6$Sn$_6$, we briefly remark on the possible surface terminations of the 166 kagome materials. As shown in Fig. 1c, the unit cell of ScV$_6$Sn$_6$ consists of one ScSn$_2$ layer, one hexagonal Sn$_2$ layer, and two V$_3$Sn kagome layers; this HfFe$_6$Ge$_6$-type 166 structure can expose complex surface terminations upon cleaving. We note that previous studies on the 166 kagome materials yield inconsistent interpretations on the surface terminations (Supplementary Section 1). To resolve this issue, we performed spatially-resolved ARPES and XPS experiments on ScV$_6$Sn$_6$ using micro-focused synchrotron radiation (Fig. 1k,l). As summarized in Fig. 1m-o, we clearly identified three different surface domains characterized by dramatically different valence band structures and Sn $4d_{3/2}$, $4d_{5/2}$ core level spectra (D1, D2, and D3 domains, respectively). By comparing the ARPES and XPS spectra at each domain to the slab DFT calculations of various geometries, we unambiguously assigned the D1, D2, and D3 domains to the ScSn$_2$, V$_3$Sn, and Sn$_2$ surface terminations, respectively (Supplementary Section 2,3). Below we focus on the results obtained on the V$_3$Sn termination (D2), which best represents the bulk electronic structure of ScV$_6$Sn$_6$ based on the slab calculation (Supplementary Section 3).

Figure 2 displays our analysis of the low-energy electronic structure of ScV$_6$Sn$_6$. Similar to the case of $A$V$_3$Sb$_5$, we identified multiple kagome-derived vHSs near the Fermi level. In Fig. 2a-c, we present a three-dimensional stack of the ARPES spectra measured at the vicinity of M point (see the momentum positions of the cut 1-4 in Fig. 2j). From these plots, one can comprehensively understand the dispersions along both the Γ-M-Γ (see solid lines in the cut 4) and the K-M-K direction (see dashed lines across the cut 1 to 4). As shown in Fig. 2a,b, we identified two bands having electron-like character along the K-M-K direction and hole-like character along the Γ-M-Γ direction; these bands thus form saddle point structures or vHSs at the M point as predicted from the model kagome lattice dispersion (see also schematics in Fig. 2g,h). As shown in Fig. 2c, we also observed one additional vHS with inverted concavity, i.e., hole-like dispersion along the K-M-K and electron-like dispersion along the Γ-M-Γ direction (see schematics in Fig. 2i). The density functional theory (DFT) calculations in Fig. 2d-f closely reproduce the experimental results, revealing that the three vHSs in Fig. 2a-c respectively originate from the $d_{xy}$, $d_{xz}$, and $d_{z2}$ local orbital degrees of freedom in the V kagome lattice (Supplementary Section 4). In the kagome lattice, the sublattice character of vHS – *pure* (*p*) or *mixed* (*m*) sublattice character – is also a topic of great interest, which critically determines the relevance of the on-site and long-range Coulomb interactions and the leading electronic instabilities.[3,4,23] By analyzing the sublattice weight distribution near the M point, we revealed that all three vHSs in ScV$_6$Sn$_6$ are *p*-type vHs having pure sublattice character (Supplementary Section 5). In sum, our analysis provides the complete characterizations of the dispersions, orbital characters, and sublattice types of the vHSs in ScV$_6$Sn$_6$.

Notably, the $d_{xy}$ and $d_{z^2}$ vHS of ScV$_6$Sn$_6$ locate very close to the Fermi level at $-0.02 \pm 0.01$ eV and $-0.03 \pm 0.02$ (Fig. 2a,c), while the $d_{xz}$ vHS is positioned at higher binding energy $\approx -0.40 \pm 0.03$ eV (Fig. 2b). The former vHSs contribute to the diverging density of states at the Fermi level and can in principle promote various electronic instabilities including the charge orderings. This scenario indeed applies to the case of *A*V$_3$Sb$_5$, where the vHSs at the Fermi level develop charge order gaps and directly contribute to the stabilization of the 2×2 charge order.[21,22,38,39] To test this scenario in ScV$_6$Sn$_6$, we tracked the temperature evolution of the vHSs across the $\sqrt{3}\times\sqrt{3}$ charge order transition. Comparing the ARPES dispersions in the normal (Fig. 2k,m) and charge-ordered states (Fig. 2l,n), we observed that all vHSs in ScV$_6$Sn$_6$ stay surprisingly unaltered across $T_{CO}$, despite the fact that the vHS momentum (i.e., M point) lies at the folded Brillouin zone boundary of the $\sqrt{3}\times\sqrt{3}$ charge order (see the schematics in Fig. 2j). This observation indicates that in stark contrast to the case of *A*V$_3$Sb$_5$, the intrinsic electronic instability of the V kagome lattice plays a marginal role in driving the $\sqrt{3}\times\sqrt{3}$ charge order in ScV$_6$Sn$_6$.

After ruling out the vHS, we explore the electronic structure of ScV$_6$Sn$_6$ in the extended momentum range (Fig. 3), to identify the bands actually relevant to the $\sqrt{3}\times\sqrt{3}$ charge order transition. Fig. 3a,b display the Fermi surfaces of ScV$_6$Sn$_6$ measured at the normal and charge-ordered state, respectively. The major reconstruction of the Fermi surface across $T_{CO}$ is apparent from our data: the circular intensity pattern centered at $\bar{\Gamma}$ in the normal state is modified to the star-shaped pattern in the charge-ordered state as highlighted with the cyan and orange guidelines. To better understand this change, we also present the corresponding energy-momentum dispersions along the $\bar{\Gamma}$-$\bar{M}$ direction in Fig. 3e,f. In the normal state dispersion (Fig. 3e), we observe a large electron pocket centered at $\bar{\Gamma}$, which constructs the circular intensity pattern in the normal state Fermi surface (Fig. 3a). Below $T_{CO}$ (Fig. 3f), this electron band bends toward the higher binding energy and develops a substantial charge order gap at the Fermi level. This opening of the charge order gap depletes the intensity in the Fermi surface along the $\bar{\Gamma}$-$\bar{M}$ direction and explains the star-shaped Fermi surface observed in the charge-ordered state (Fig. 3b). The momentum position of the charge order gap is at about two-thirds of the $\bar{\Gamma}$-$\bar{M}$ direction, which excellently matches with the folded Brillouin zone boundary of the $\sqrt{3}\times\sqrt{3}$ phase (see Fig. 2j). We note that the band renormalization and charge order gap is also observed in other surface terminations supporting their bulk origin (see Supplementary Fig. S7 for the charge order gaps measured in the D1 termination). As shown in Fig 3g,h, our DFT calculations closely capture the experimental results, reproducing the large electron pocket at $\bar{\Gamma}$ in the normal state (Fig. 3g) and opening of the charge order gap $\Delta_{CO} \approx 260$ meV across the Fermi level in the charge order state (Fig. 3h). Notably, the magnitude of the charge order gap in ScV$_6$Sn$_6$ is significantly larger than the $\Delta_{CO} \approx 80$ meV of $A$V$_3$Sb$_5$ despite the comparable $T_{CO}$ in two systems.[21,22]

Importantly, this large electron pocket at $\bar{\Gamma}$, which is closely tied to the $\sqrt{3}\times\sqrt{3}$ charge order, has dominant planar Sn character (i.e., Sn(1) in Fig. 1c). To illustrate this, we present the DFT band structure of ScV$_6$Sn$_6$ in Fig. 4a, along with the V and Sn(1) orbital-projected calculations in Fig. 4b,c. The corresponding Fermi surfaces are also shown in the insets. In the V orbital-projected calculation (Fig. 4b), multiple Dirac bands at $\bar{K}$ and van Hove singularities at $\bar{M}$ originating from the V kagome lattice can be clearly identified. Overall, the V spectral weights dominate the Fermi surface near the zone boundary. In contrast, the Fermi surface near the zone center $\bar{\Gamma}$ has dominant Sn(1) orbital character, as shown in the inset of Fig. 4c. We emphasize that it is this Sn(1) band at $\bar{\Gamma}$ that develops the charge order gap and reconstructs the Fermi surface across $T_{CO}$ (Fig. 3), while the V kagome bands near $\bar{K}$ and $\bar{M}$ remain unaltered across $T_{CO}$ (Fig. 2). Our results thus highlight that the $\sqrt{3}\times\sqrt{3}$ charge order of ScV$_6$Sn$_6$ is tied to the structural components other than the V kagome lattice, especially to the planar Sn atoms.

The above conclusion from the electronic sector is further supported by our phonon mode calculations presented in Fig. 4d-g. As shown in Fig. 4d, the phonon dispersions of ScV$_6$Sn$_6$ display the continuum of unstable phonon modes centered at $H$, consistent with the $\sqrt{3}\times\sqrt{3}$ reconstruction in ScV$_6$Sn$_6$ at low temperature. By projecting the phonon density of states to the Sc, V, and Sn(1), Sn(2), Sn(3) sites in the unit cell, we revealed that the unstable phonon modes are associated with the structural distortions involving the planar Sn(1) and Sc sites, while the contribution from the V kagome layer is negligible (Fig. 4e). This result not only explains our observation of the large charge order gap on the Sn(1) bands (Fig. 3) and the marginal change of the V kagome bands (Fig. 2), but also is fully consistent with the X-ray refined charge order structure of ScV$_6$Sn$_6$ that revealed the dominant distortions in the Sn(1) and Sc sites.[37] It is also instructive to compare the phonon modes of ScV$_6$Sn$_6$ to those of the CsV$_3$Sb$_5$ shown in Fig. 4f,g. In stark contrast to the case of ScV$_6$Sn$_6$, the unstable phonon modes of CsV$_3$Sb$_5$ at $M$ and $L$ (associated with the 2×2 charge order) accompany the dominant displacement of the V atoms, and reflect the intrinsic electronic instability from the V kagome layers (Supplementary Fig. S10).[40]

Finally, we construct the phase diagram of Sc(V$_{1-x}$Cr$_x$)$_6$Sn$_6$ series to understand the evolution of charge order with carrier doping (Fig. 5). The charge order phase remains robust in the wide-doping range, up to doping $x \approx 0.10$ charges per V atom. In the framework of the virtual crystal approximation (Supplementary Fig. S11), this indicates that the charge order phase remains stable up to the order of 120 meV Fermi level shift, further making the electronic scenarios sensitive to the Fermi level, such as the Fermi surface or vHS nesting, unlikely. We note that the response of the charge order to carrier doping is again highly different in CsV$_3$Sb$_5$, where the charge order rapidly vanishes after the $x \approx 0.02 \sim 0.03$ charge doping per V atom, regardless of the doping methods.[31,41]

In summary, the present work elucidates the origin of the $\sqrt{3} \times \sqrt{3}$ charge order in the newly discovered kagome metal ScV$_6$Sn$_6$. Our comprehensive characterizations of the electronic structure, phonon dispersion, and phase diagram coherently emphasize the essential role of the structural degrees of freedom other than the V kagome lattice in driving the charge order. In this context, the nature of the $\sqrt{3} \times \sqrt{3}$ charge order in ScV$_6$Sn$_6$ is fundamentally different from the 2×2 charge order in the archetype kagome metal $A$V$_3$Sb$_5$, where the electronic instability in the V kagome lattice plays a major role. As discussed in the introduction, the true charge disproportionation phases in the kagome lattice can support the exotica of physics, including the fractionalization of elementary particles. Our study thus emphasizes that the search for new kagome quantum materials hosting various types of genuine charge orders should be continued.

## Methods

**Single crystal synthesis and characterization.**

Single crystals of $ScV_6Sn_6$ and $Sc(V_{1-x}Cr_x)_6Sn_6$ doping series were grown by typical self-flux methods. Scandium pieces (99.9 % Research Chemicals), Vanadium pieces (99.7 % Alfa Aeser), and Sn ingot (99.99 % Alfa Aeser) were put in the Alumina crucible with frit disc, then sealed in Ar-gas purged evacuated quartz tube. Ampule was heated at 1100 °C for 24 hrs, then slow cooled to 800 °C with 1~2°C/hr cooling ratio. To remove the flux, ampule was centrifuged at 800 °C. V and Cr ratios of the doping series were confirmed using energy dispersive spectroscopy. Electrical Resistivity measurements was performed with Physical Properties Measurement System (PPMS, Quantum design) using a conventional 4 probe method. The X-ray diffraction measurements were conducted using Cu $K_{\alpha 1}$ source ($\lambda$ = 1.54 Å) and 6-axis diffractometer. We identified (1/3 1/3 19/3), (1/3 1/3 20/3), (2/3 2/3 19/3), and (2/3 2/3 20/3) peaks associated with the $\sqrt{3}\times\sqrt{3}\times 3$ charge order, all displaying the same temperature-dependence.

**ARPES experiments.** ARPES experiments were conducted at Beamline 7.0.2 (MAESTRO) and Beamline 4.0.3 (MERLIN) of the Advanced Light Source, equipped with R4000 and R8000 hemispherical electron analysers (Scienta Omicron), respectively. The samples were cleaved inside ultra-high vacuum chambers with a base pressure better than 4 x $10^{-11}$ Torr. To identify high-symmetry planes of the three-dimensional bulk Brillouin zone of $ScV_6Sn_6$, the photon energy dependent ARPES measurements were performed in a wide phonon energy range from 60 eV to 200 eV. By comparing the experimental $k_z$ dispersion to the DFT band structure, we identified 129 eV and 115 eV photon energies measuring the $k_z \approx 0$ and $k_z \approx \pi$ high-symmetry planes, respectively. The data in Fig. 2 are acquired with 129 eV photons, while the data in Fig. 3 are measured with 115 eV photons. All normal state (charge-ordered state) data in the main text is obtained at 120 K (6 K) using linear horizontal light polarization, unless specified. We refer to Supplementary Fig. S8 for the data measured in finer temperature steps.

**Spatially-resolved ARPES and XPS experiments.** The real-space mappings of the valence band structure and core level spectra were conducted at Beamline 7.0.2 (MAESTRO) of the Advanced Light Source. To resolve the complex surface domains of $ScV_6Sn_6$, we used the micro-focused synchrotron of lateral dimension 30 × 30 $\mu m^2$. The domain dependent ARPES and XPS spectra are compared to the slab DFT calculations of various geometries, to assign the atomic termination layer to each domain (Supplementary Section 2,3).

**DFT calculations.** DFT calculations were performed using the Vienna AB initio Simulation Package software.[42,43] The generalized-gradient approximation Perdew-Burke-Ernzerhof exchange-correlation functional was chosen to calculate the exchange-correlation energy.[44] The pseudopotential was defined based on the projector augmented-wave method.[45] VASPKIT software was used for pre- and post-processing of DFT calculated information.[46] For the bulk band calculation of $ScV_6Sn_6$, we used the lattice parameters $(a, b, c)$ = (5.456 Å, 5.456 Å, 9.230 Å) which is obtained by relaxing the reported single crystal refinement data.[37] Relaxation is performed at the 350 eV kinetic energy cutoff that fully covers the atomic energy. The static electronic structure was calculated using a $\Gamma$-centered $k$-point mesh, 15x15x8 for the normal state structure and 8x8x3 for the charge ordered structure. We present the overall band dispersion of $ScV_6Sn_6$ in the normal state with and without spin-orbit coupling in Supplementary Fig. S9a. Supplementary Fig. S9b displays the unfolded DFT band dispersion in the charge ordered state. To understand the termination-dependence of the valence band and core level spectra, we performed the slab DFT calculation on all possible charge neutral slab configurations of $ScV_6Sn_6$ relaxed at the 350 eV kinetic cutoff energy (Supplementary Section 2,3). $\Gamma$-centered 11x11x1 $k$-point mesh were used for the slab band calculation. Each slab has 20 atomic layers and the vacuum was fixed at 20 Å.

**DFPT calculations.** Phonon dispersions were computed within the density functional perturbation theory (DFPT) framework. Input parameters were generated from the 3x3x2 supercell using Phonopy software.[47,48] We compared the phonon modes of two kagome metals, $ScV_6Sn_6$ and $CsV_3Sb_5$. Various smearing factors 0.10, 0.125, 0.15, 0.175, and 0.20 were tested for both compounds. We present the smearing factor-dependent phonon dispersions in Supplementary Fig. S10.

**Acknowledgements**

The works at Max Planck POSTECH/Korea Research Initiative were supported by the National Research Foundation of Korea funded by the Ministry of Science and ICT, Grant No. 2022M3H4A1A04074153 and 2020M3H4A2084417. This research used resources of the Advanced Light Source, a U.S. DOE Office of Science User Facility under contract no. DE-AC02-05CH11231. R.C. acknowledges the support from the Air Force Office of Scientific Research grant FA9550-22-1-043, and the STC Center for Integrated Quantum Materials, NSF grant DMR-1231319. M.K. acknowledges a Samsung Scholarship from the Samsung Foundation of Culture. This paper is supported by Basic Science Research Institute Fund, whose NRF grant number is 2021R1A6A1A10042944.

**Author contributions**

R.C., M.K., and J.-H.P. conceived the project; S.L., J.K., J.Y., and S.P. performed the ARPES experiments and analyzed the resulting data with help from J.D., C.J., A.B., and E.L.; S.L. performed the theoretical calculations; C.W. synthesized and characterized the crystals.; S.L. and M.K. wrote the manuscript with input from all coauthors.

**Data availability**

The datasets presented within this study are available from the corresponding authors upon reasonable request.

**Competing interests**

The authors declare no competing interests

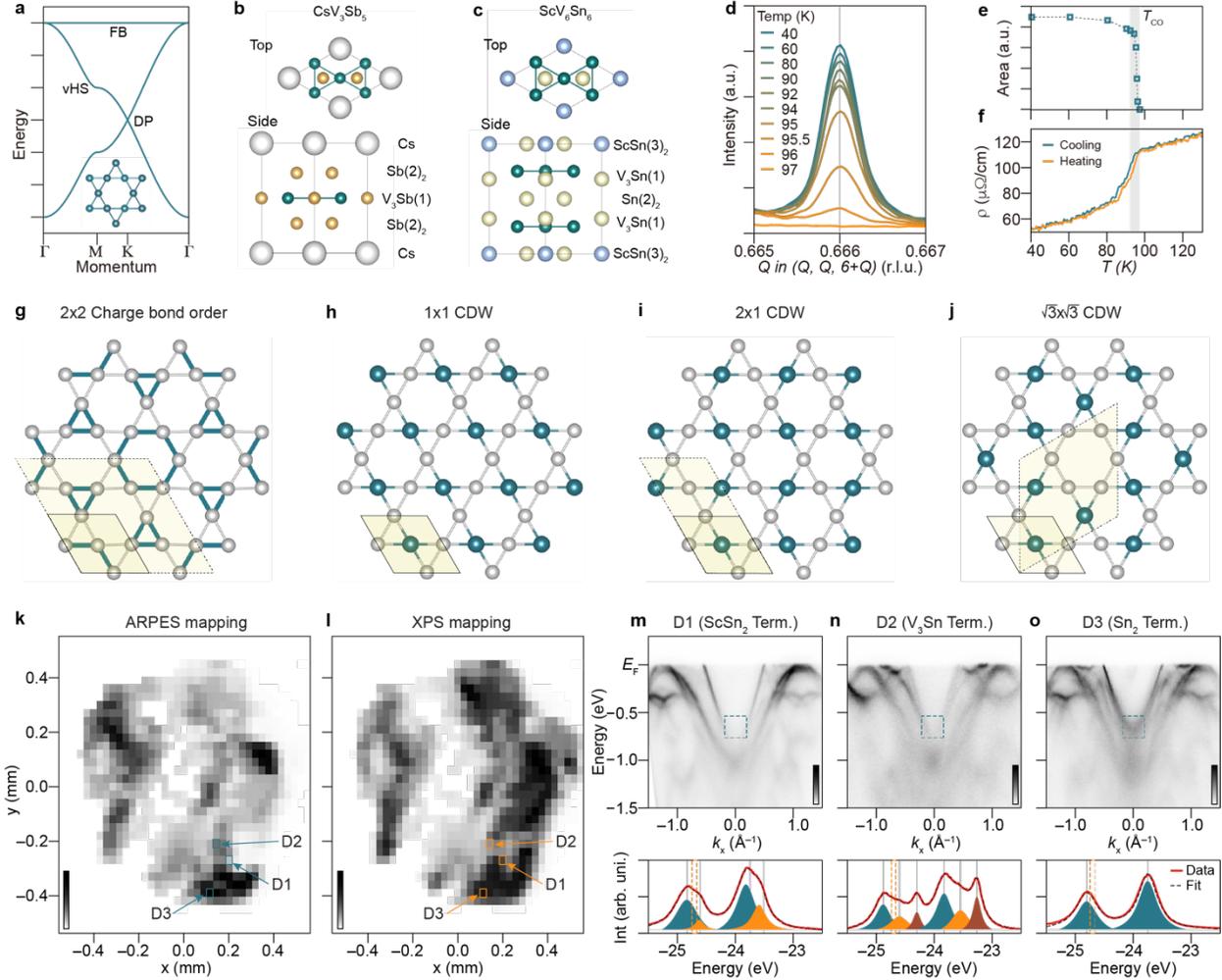

**Figure 1 | Novel charge orders in kagome metal. a,** Prototypical electronic structure of the kagome lattice featuring the Dirac point, vHS, and flat band. Inset displays the lattice structure. **b,c,** Crystal structure of kagome metals $CsV_3Sb_5$ (a) and $ScV_6Sn_6$ (b) sharing the same V kagome plane. **d,e,** Temperature-dependent X-ray diffraction profile and integrated peak area of the (2/3, 2/3, 20/3) charge order peak in $ScV_6Sn_6$, respectively. **f,** Temperature-dependence of the in-plane resistivity of $ScV_6Sn_6$ around the charge order transition. **g-j,** Various types of charge orders predicted from the extended Hubbard model on kagome lattice. **k,l,** Real space mapping of the ARPES and Sn 4$d$ XPS intensity of the $ScV_6Sn_6$ sample. Three different surface domains (D1, D2, and D3) with dramatically different ARPES and XPS spectra were identified. The representative D1, D2, and D3 domain positions are marked in k,l. **m-o,** ARPES (Top panel) and XPS (bottom panel) spectra of $ScV_6Sn_6$ on D1, D2, and D3 domains, respectively. The dashed cyan and orange boxes in m-o represent the area where the ARPES and XPS intensities are integrated and plotted in k,l.

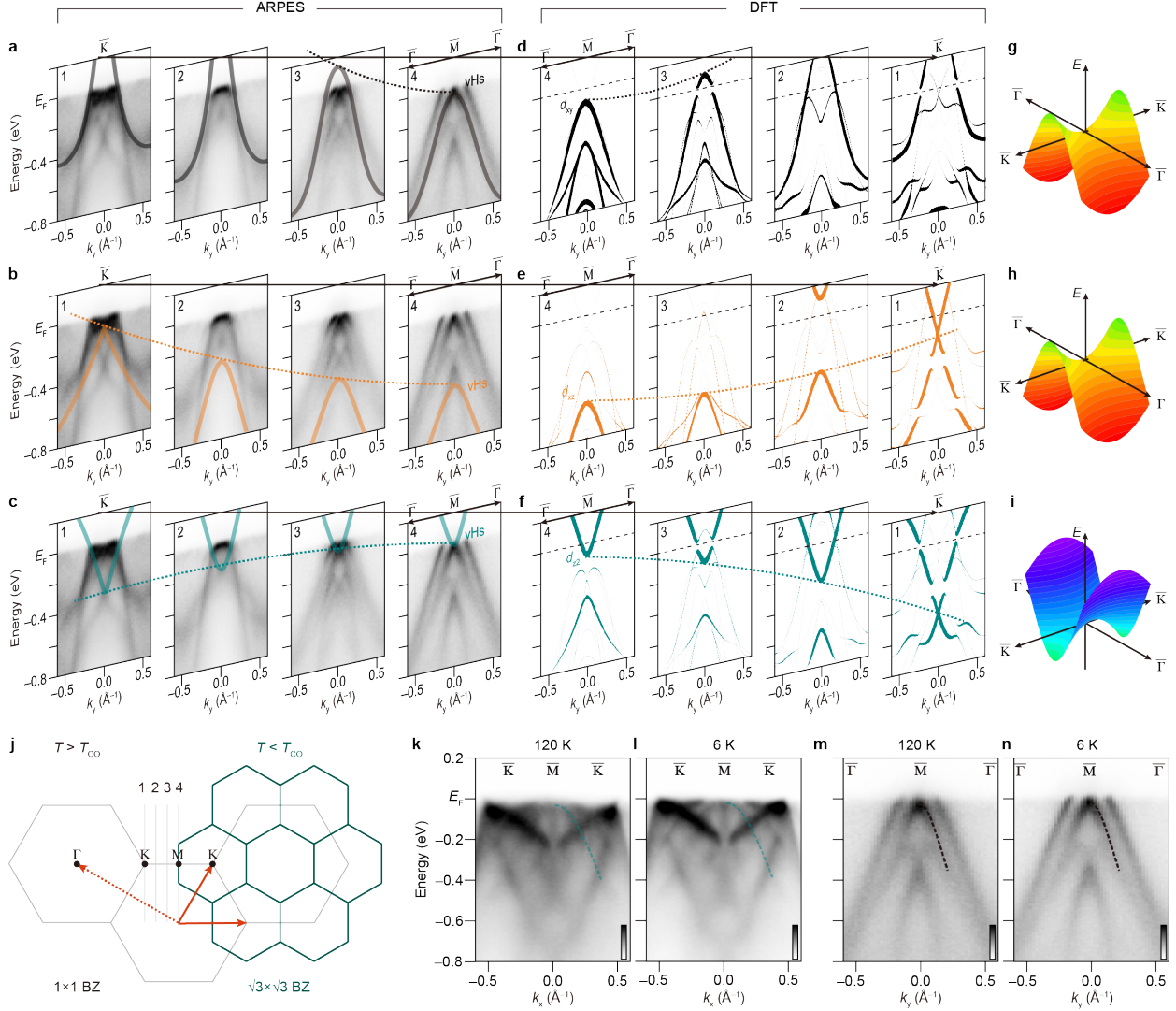

**Figure 2 | Characterization of the van Hove singularities in ScV$_6$Sn$_6$. a-c,** Experimentally identified vHS dispersions in ScV$_6$Sn$_6$ using ARPES. The cut 1-4 in a-c plot the ARPES spectra measured perpendicular to the K-M-K direction, with the cut 1 crossing the K point and the cut 4 crossing the M point (see panel j for the exact momentum positions of cut 1-4). We identified three coexisting vHS in ScV$_6$Sn$_6$ as marked with black, orange, and cyan guidelines in a-c, respectively. **d-e,** Corresponding DFT band structures of ScV$_6$Sn$_6$ for comparison with the ARPES spectra in a-c. The fat bands in d-f represent the spectral weight of the $d_{xy}$, $d_{xz}$, and $d_{z2}$ local orbitals, respectively. **g-i,** Schematics of the saddle point dispersions or vHSs. The concavity of vHS is identical for $d_{xy}$ and $d_{xz}$ vHS, while it becomes opposite for the $d_{z2}$ vHS. **j,** Schematics of the pristine (grey hexagons) and the √3×√3 folded (cyan hexagons) in-plane Brillouin zones of the ScV$_6$Sn$_6$. Dashed and solid orange arrows represent the reciprocal lattice vectors. **k-n,** Temperature dependence of the vHSs across the charge order transition. The cyan and black dashed lines are guide for the eye for the $d_{z2}$ and $d_{xy}$ vHS dispersions near the Fermi level, respectively. All data were collected with 129 eV photons, measuring the $k_z \approx 0$ high-symmetry plane of the three-dimensional Brillouin zone.

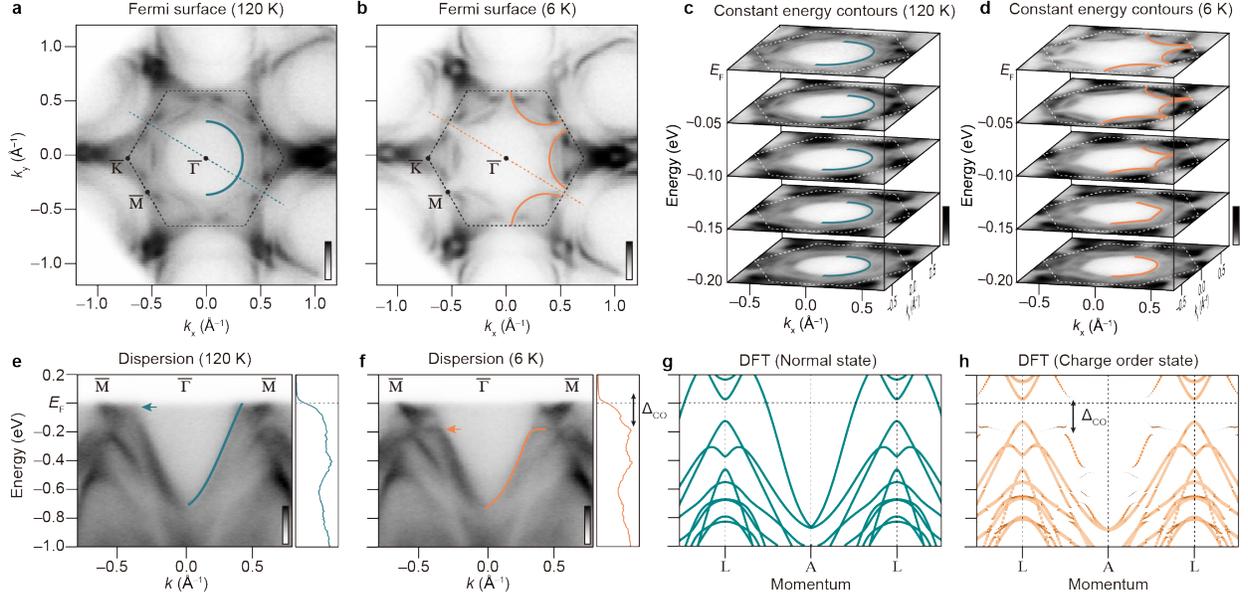

**Figure 3 | Fermi surface reconstruction and charge order gap opening across the $\sqrt{3}\times\sqrt{3}$ transition. a,b,** The Fermi surface of ScV$_6$Sn$_6$ in the normal and charge-ordered state, respectively. The data were obtained using 115 eV photons corresponding to the $k_z \approx \pi$ high-symmetry plane. The cyan and orange solid lines in a-f are guide for the eye highlighting the band dispersion around the $\bar{\Gamma}$ point. **c,d,** Stack of the constant energy contours of ScV$_6$Sn$_6$ in the normal and charge-ordered state, respectively. **e,f,** Normal and charge-ordered state dispersion of ScV$_6$Sn$_6$ measured along the $\bar{\Gamma}$-$\bar{M}$ high symmetry direction marked in a,b. The cyan and orange arrows highlight the back bending of the dispersion due to the charge order gap opening at the Fermi level. The right panels in e,f display the energy distribution curves measured at the $\sqrt{3}\times\sqrt{3}$ charge order Brillouin zone boundary, i.e. at the two thirds of $\bar{\Gamma}$-$\bar{M}$ momentum. The spectral weight shift and charge order gap opening is evident from the energy distribution curves as marked with the black arrow in f. **g,h,** DFT band structure of ScV$_6$Sn$_6$ in the normal and charge-ordered state, respectively. The band structure in the charge-ordered state is unfolded to the pristine Brillouin zone to facilitate comparison. The black arrow in h indicates the charge order gap observed in the experiment.

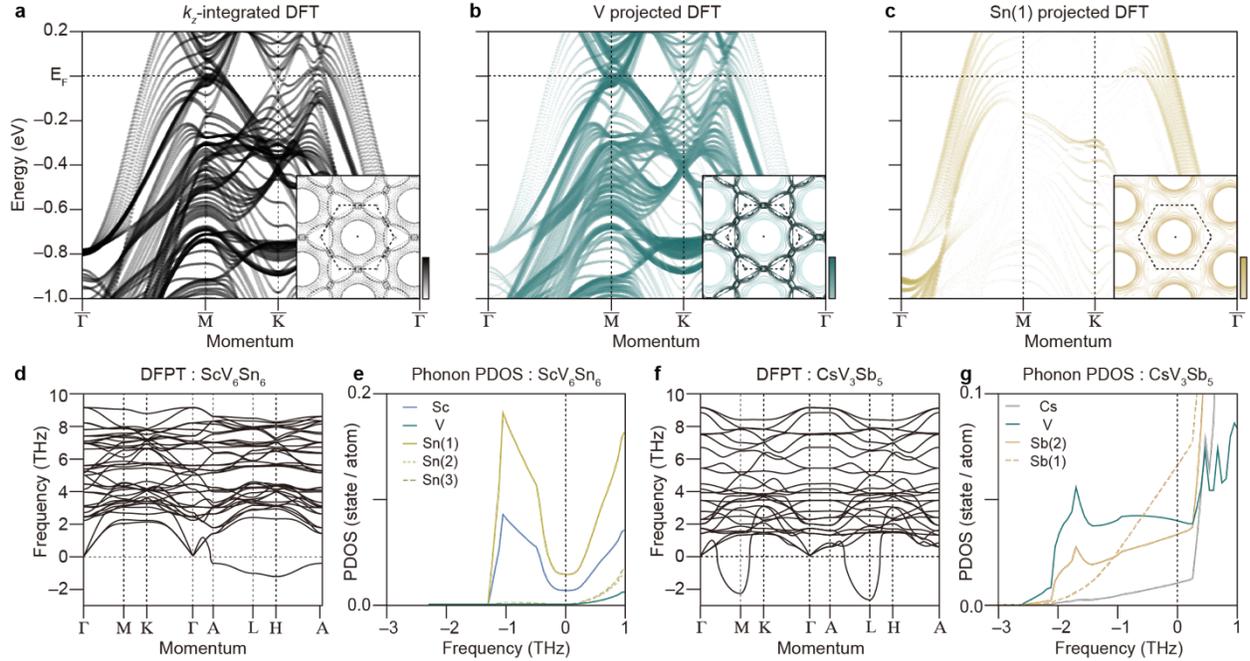

**Figure 4 | Element- and site-resolved DFT band structure and DFPT phonon modes of ScV$_6$Sn$_6$. a-c,** $k_z$-integrated DFT band structure and its projection to the V and Sn(1) orbitals, respectively. The insets display the corresponding Fermi surfaces. **d,** The phonon dispersion of ScV$_6$Sn$_6$ obtained from the DFPT calculation. **e,** The phonon partial density of states of ScV$_6$Sn$_6$ projected to the Sc, V, Sn(1), Sn(2), and Sn(3) sites of the unit cell (see Fig. 1c). **f,g,** The DFPT phonon dispersion and phonon partial density of states of CsV$_3$Sb$_5$, respectively, for comparison with the ScV$_6$Sn$_6$ in d,e.

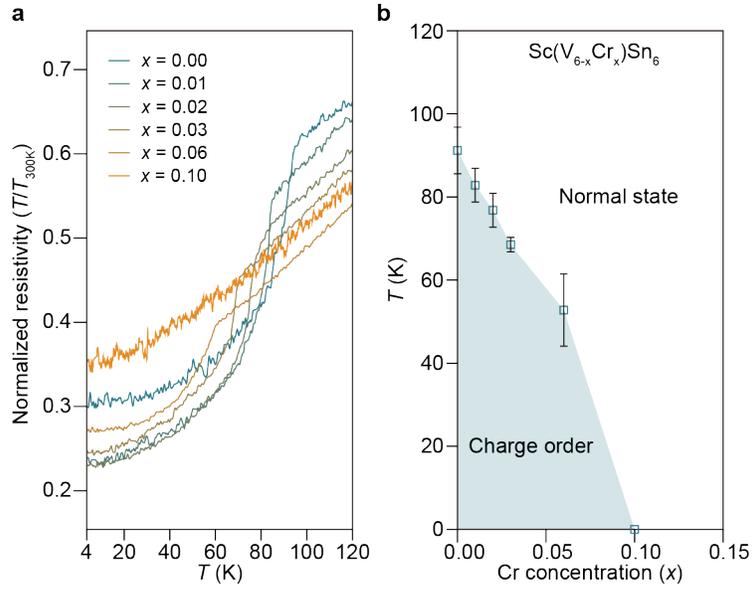

**Figure 5 | Phase diagram of charge order in Sc(V$_{1-x}$Cr$_x$)$_6$Sn$_6$ series. a,** Evolution of the normalized resistivity $T/T_{300K}$ as a function of Cr-doping in $x$ = 0, 0.01, 0.02, 0.03, and 0.06 samples. **b,** Doping-temperature phase diagram of the charge order in ScV$_6$Sn$_6$.